# Testing in Global Software Development – A Pattern Approach

Anneke Pehmöller
Capgemini
Munich, Germany

Frank Salger
Capgemini
Munich, Germany

Stefan Wagner
University of Stuttgart
Stuttgart, Germany
stefan.wagner@
informatik.uni-stuttgart.de

**Abstract**

Although testing is critical in GSD, its application in this context has not been deeply investigated so far. This work investigates testing in GSD. It provides support for test managers acting in a globally distributed environment. With this it closes a gap. The leading question is "What problems exist in testing in GSD and how can they be addressed in projects?" Decomposing this question we a) identify problems of testing in GSD projects and b) provide good practices to support practitioners in testing in GSD projects. The research is realized in the context of Capgemini Germany. Our contribution to solving the stated research problem is a collection of 16 patterns for testing in GSD projects. For practitioners the usage of the patterns is simplified by various views on the patterns. Herewith we stipulate research and support project managers and test managers in the realization of testing in GSD projects.

## 1. Introduction

Global Software Development (GSD) is a widely-used approach in software development as the industry reaches for benefits such as low labor costs, access to a large workforce, and time-zone effectiveness [1, 2]. Compared to co-located software development projects, GSD projects face the additional characteristics geographic, temporal, cultural, and technological distance (see, e.g., [1], [3-7]). Global distribution and its implications on software development has been investigated intensively in recent years. But the impact of distribution on testing is one field that has not been in the focus yet. However, the need to investigate test in the context of GSD challenges has been identified already by several researchers (see, e.g., [8], [1-2], [9-11]). Also the industrial community reveals the need for investigation and solutions in this field. Experience at Capgemini shows that testing in GSD is a little investigated issue that needs practicable solutions in the context of custom software development. The company-own Offshore Custom Software Development (OCSD) method [12] does address issues such as project management, handover points, and quality assessments (see, e.g., [12-14]) but is sparse on testing.

**Research Problem:** It is not yet clear what impact the challenges of GSD have on single testing activities and test management. Only selected aspects of testing in GSD have been investigated and solutions proposed. With this work we contribute to closing the described gap regarding testing in GSD projects. The leading research question we ask is: "What problems exist in testing in GSD and how can they be addressed in projects?"

**Research Objective:** By investigating resources from literature and industrial custom software development (CSD) projects we aim to answer the above mentioned research questions. We focus on two objectives: First, we want to present problems that result from the impact of challenges of GSD on testing. And second, we aim to extract practiced solutions to support test managers and project managers by providing reusable best practices for testing in GSD projects.

**Contribution**: With this work we contribute to the research area of testing in GSD. We provide practicable support for test managers and project managers who meet the challenge to set-up and successfully realize testing in a GSD and CSD context. We do so by providing patterns for testing in GSD projects. Each pattern describes a concrete problem and provides a solution to solve this problem.

**Context**: This work resulted from a research cooperation between the Technische Universität München (TUM) in cooperation with Capgemini Germany. The goal is to investigate the topic of testing in GSD projects. Therefore, we explore relevant literature and interview test managers and project managers of real-life projects.

**Outline:** The structure of this paper follows the guidelines for reporting case study research in software engineering proposed in [15]. In section 2 we provide background information on GSD in general and on related work on testing in GSD in specific. In section 3 we describe how we approached the above defined research question, how we collected data, and the analysis procedure. In section 4 we present patterns for testing in GSD as our result for the research questions. Section 5 contains our conclusions including further research options.

## 2. Related Work

Available GSD- and test-specific studies were investigated in a systematic literature review following the guidelines for performing systematic literature reviews in software engineering presented in [16]. The *data source* used for the review was Google Scholar [17]. This meta search engine searches well-known and accepted scientific databases including IEEE Xplore, ACM Digital Library, citeseer, Wiley Interscience, and Sciencedirect for references. We defined the *search string* as ((test OR "testing" OR "Test management" OR "test process") AND ("Global software development" OR "Distributed Software Development" OR "Global Software Engineering")). The search was executed using the configuration "at least summaries" (in contrast to "include citations") and not restricting the years of publications. For the selection of studies, we define the following mandatory *inclusion criteria*: a) The resource is in English or German. b) The title contains "test" (also accepted when in combination with other terms, e.g., test management) referring to software tests. c) The title hints to GSD, e.g., one of the terms "Global software development", "Distributed Software Development", "Global Software Engineering" or similar is mentioned. Additionally, the following *exclusion criteria* are specified: d) The study has the topic of "remote testing". Remote testing refers to testing of physically distant systems. This setting might occur in GSD projects but we consider it not directly connected to our research questions. e) The report deals with testing of "distributed software" or "distributed systems".  This is out of scope for this work. f) The resource is a patent. We focus on report types that present scientific research, e.g., conference and journal articles. g) The resource presents only statements of future research. Therewith we focus on results of investigations. The quality of the selected resources were assessed using the following *quality criterion*: h) The study has been accomplished by at least two researchers or/and has been subject to a review, e.g., as practiced for the selection of conference contributions and journal articles. Resources that did not fulfill the criteria were not further considered.

**General GSD issues:** In [1-2] an extensive overview of literature available on issues in GSD is presented. We briefly name some of the general GSD issues. Missing informal communication (see, e.g. [1], [3], [18-19]), control [20], trust [1], [20-21], visibility [18] and knowledge management [22-23] are issues resulting mainly from geographic distance. Inefficient asynchronous communication [1], [3], [19] is caused by temporal distance in distributed teams which can also result in reduced efficiency and speed (see, e.g., [4], [21], [24-25]). Differing terminologies, roles and unsynchronized processes are issues of organizational culture (see, e.g., [5], [19], [25-26]). Difficulties can also be caused by different languages [21] and communication behavior [21], [27] are issues of national culture. Technological distance concerns differing technological equipment, format standards, and infrastructural issues as low band width and connectivity problems [1], [4-5], [19]. All these challenges (probably) somehow influence testing activities in a distributed setting. Nevertheless, they do not address the specific research problem.

**Testing in GSD:** In [11] the question "How can software testing in a globally distributed virtual team environment be effectively carried out?" is investigated. [28] propose Test-driven Global Software Development as a solution to keeping knowledge on requirements and their changes consistent at all involved sites. With their tool Global Access Testing Environment (GATE) [9] provide a solution to manage test-related information and thus support effective testing in GSD. In [8] the authors propose a workflow-based tool for the management of the test process. Model-based testing in GSD has been investigated in [29]. Focusing on offshoring test automation, [30] investigates three cases of outsourcing activities of test automation to offshore providers. In [31] experiences on training a performance test team located in India by a performance engineer in New Jersey (USA) are reported.

The described findings and proposed solutions are helpful contributions. Still, a structured, comprehensive presentation of problems and solutions is missing. We do see the need for more investigation on how the challenges of GSD impact testing, what consequences this has and what practicable solutions can be proposed.

## 3. Case Study Design

We use an exploratory case study approach with the goal to investigate contemporary phenomena in their context [15]. With our research we investigate qualitative data, i.e., data collected directly from software development projects. We base our research on the idea of grounded theory proposed in [32] and thus use a hypothesis generating technique.

**Research questions:** The leading *research question* we ask is "What problems exist in testing in GSD and how can they be addressed in projects?" Decomposing this question we investigate the following problems. a) GSD projects differ from co-located projects: What challenges exist? b) We expect that these challenges impact the development life cycle: What impact do these challenges have on testing? What implications do these impacts have for the test organization, the test activities, and test artifacts? c) Solutions might have been found already: How can the impact of challenges of GSD on testing be addressed? What best practices regarding testing have been used in GSD projects?

**Case and subjects selection:** Our research bases on qualitative data collected at the industrial partner Capgemini Germany (for more information see http://www.de.capgemini.com). For a couple of years, Capgemini Germany has realized many projects with globally distributed teams. Besides Germany, sites in Poland and India are involved. Activities are distributed within the Capgemini group and thus an intra-organizational approach is realized. The company-own Offshore Custom Software Development (OCSD) method provides guidance and best practices for GSD projects. At Capgemini Germany, GSD projects follow a so called "one team approach". That is the complete integration of all team members irrespective of the site they are working at. All colleagues are full members of one team. Still the main responsibility of the project, namely the project management is kept in Germany. Regarding the development process, projects at Capgemini sd\&m have to be distinguished as various process models from almost strict waterfall to iterative to agile methods are accomplished. Organization and distribution in the GSD context is individual for all projects. This is characteristically of the domain of custom software development. All study subjects and objects are from the described context. *Study objects* are four projects. The projects were selected by a GSD expert who knew that they are GSD projects. All projects were so called offshore projects with German and Indian team members. As *study subjects* we focused on test managers and project managers as interview partners. All of them are located at the German site. We took five study subjects into account, two project managers (PM) and three test managers (TM) of which one TM also took on the role of PM.

**Data collection procedures:** The collection of practical data included mainly interviews with project experts. The collection of this data was accomplished within a time frame of twelve weeks. The source for our qualitative data are the above mentioned study subjects and objects. As preparation of the *interviews with project experts*, the interviewee filled out a short questionnaire of three pages. This gave a first impression of the project's organization. Secondly, the interview was held by telephone with one interviewee at a time for approximately one hour. The conversation was recorded. Interviews were semi-structured. The interview was kept flexible to react to the flow of the conversation and information given by the interviewee. Transcripts of all interviews were made for later analysis.

**Analysis procedure:** For the analysis of practical data collected during project interviews, we used a grounded theory approach. We generated a theory from qualitative data. During the first analysis of the interviews' transcripts, we developed the idea of extracting patterns to provide a solution for our research problem. Patterns are an established and well-known format to present knowledge of software engineering [33] (on patterns see also [34-35]). Patterns have the benefits to be lightweight, easily extendable and focused. We used a structured process with the goal to extract patterns from the available data. For each project we analyzed the transcript of the interview. Test-specific and GSD-specific problems and in the project realized approaches and solutions are identified and excerpted. The found problems and approaches serve as basis to formulate ideas for patterns which are in the first step shortly described with a title. The result is a list of data records of test-specific problems, approaches, and ideas for patterns for each project. Analyzing all data records, patterns were identified and described in detail. For each pattern it is evaluated if it is a "good" pattern. Based on [33], [35], [37-38] we defined a list of criteria a good pattern has to fulfill. A good pattern is relevant, encapsulated, practiced, applicable/ practical and comprehensible.

## 4. Results

The analysis procedure resulted in 16 patterns. Table 1 shows a problem/solution summary (as proposed in the pattern Problem/Solution summary, see [39]) of all patterns.

**Table 1. Problem/ Solution summary with highlighted patterns that are described in the following.**

| Problem | Solution | Pattern |
|---|---|---|
| How do you control that the knowledge transfer from On-specification to Off-test is efficient enough to ensure the knowledge needed to fulfill desired level of (specification) coverage by the Off-test specification? | After knowledge transfer from the specification team the Off-testers write test cases. The On-expert reviews these to check if the knowledge transfer was successful. | Test cases as MoU* of knowledge transfer |
| How do you prove functioning of the system when you assign complete system test to Off-test team but the final responsibility for the product is yours (On)? | After Off-system test an On-expert executes a pre-acceptance test using an exploratory/experience-based testing technique. | Light-weight pre-acceptance test |
| How do you effectively communicate bugs from On-tester to Off-developers during test/bug fixing phase? | Set-up the communication channel to be from On-TM to Off-TM or Off-Developer lead. | Communication on eye level |
| How do you ensure clear communication during test/bug fixing phase when the test team is distant to the bug fixing team? | Agree upon elements of a bug description and the bug process with all involved and use a bug tracking tool. | Use tool for bug tracking |
| How do you ensure necessary knowledge transfer to the Off-test team when the specification takes place Onshore? | During On-specification relocate Off-BAs as full team members. After specification phase, the Off-BAs move back to the Off-site and transfer the knowledge over. | Moving On-/Off-BA |
| How do you ensure that a holistic test is accomplished when different testing attitudes show in the test teams at different sites? | Check if On- and Off-team have differing testing attitudes, e.g., mainly positive test cases. If yes, assign each team the responsibility to accomplish the test for ist "favorite" testing attitude. | Complementing testing attitudes |
| How do you avoid misunderstandings between On- and Off-testers in testing? | With project start the On- and Off-TM discuss their understanding of basic test terminology and testing approaches. If they find differences in understanding, they align their understanding and define one term/process that will be used during the project. | Align understanding of general testing approach |
| How do you avoid late discovery of missing and/or wrong functionality and errors when modules are developed at distributed sites? | On the continuous integration platform set-up a (central) continuous integration test using automated integration tests. | (Central) Continuous integration test |
| How do you simplify communication between On- and Off-test team? | Set-up the test organization with an On- and Off-TM. | Mirrored TM |
| How do you set-up the testing/bug fixing process efficiently when you have a time difference between teams? | Take advantage of the time difference by defining handover points depending on the work day of involved sites. | Extension of the day |
| How do you ensure and control coverage of the specification when control is complicated by distance to the test team? | Assign each specification element an ID which the testers reference with the test cases that test this part of the specification. | Traceable test cases |
| How do you ensure test coverage (coverage of specification) when Off-testers are very specialized? | Assign a specification expert the task of review of a certain tester's field. | Tester's sparring partner |
| How do you deal with missing skills in a remote test team needed for your test strategy? | After having defined a testing strategy check if all needed skills are available in the Off-test team. If not, ensure recruitment of missing experts. | Complement test skills |
| How do you prevent misunderstandings and difficulties caused by reproduction problems of bugs in testing/bug fixing phase when the test team and the bug fixing team are geographically separated? | Set-up a test environment as master and synchronize all slaves with this configuration. | Synchronized test environments |
| How do you prevent misunderstandings and difficulties caused by reproduction problems of bugs in testing/bug fixing phase when the test team and the bug fixing team are geographically separated? | Provide one central test environment that all testers (and other team members) have full access to. | Central test environment |
| How do you provide needed test data at the Off-site under (legal) constraints? | Early in the project evaluate, if constraints exist for the usage of test data. If constraints apply, plan for production of artificial test data. | Evaluation of constraints for test data |

Pattern can be described in different styles [40]. For our needs, we define a pattern structure based on the Mandatory Pattern Elements pattern [39]. Our pattern structure contains additional but self-explaining elements to serve our needs. In the example given below, the following abbreviations are used: On = onshore (in this case referring to a site in Germany), Off = offshore (a site not in Germany, e.g., India), TM = test manager, BA = business analyst, PM = project manager.

--------------------

**Name:** Light-weight pre-acceptance test

**Problem:** How do you prove functioning of the system when you assign complete system test to Off-test team but the final responsibility for the product is yours (On)?

**Context:** a) The complete responsibility for the system test is with Off-TM and Off-test team. Thus the On-PM is the client for the result of the system test (the tested system). b) You have to decide how you check the quality of the (system-tested) product delivered by the Off-test team.

**Forces:** a) When the system test is assigned to the Off-test team, the PM might not see necessity to invest additional effort by expensive On-expert. b) Especially, when you have not worked with the Off-test team before, a check of the delivered product is inevitable.

**Solution:** a) Name an On-tester who knows the system and its critical points. This can be an On-TM or a BA who takes on the role of the On-TM. Consider the pattern *Mirrored TM*. b) After system test using a structured approach the Off-test team delivers the tested system to the On-tester. c) The On-tester executes a light-weight pre-acceptance test with an exploratory/experience-based testing method focusing on high-priority functionality. d) The On-tester reports found bugs to the Off-TM for further dealing. The pattern \textit{Use tool for bug tracking} is recommended to simplify this process. Still, communication should not only rely on the tool. e) The Off-TM takes the feedback from the pre-acceptance test as input for the approach used during structured system test (e.g. to write further test cases). Therewith, a learning effect will lead to an increase in efficiency of the system test. For this it is important to schedule needed feedback rounds, especially during first releases. The goal is to give early feedback. f) If the pre-acceptance test has reached test ending criteria (which have to be defined beforehand), the On-TM clears the system for acceptance test to the customer.

**Benefits:** a) Light-weight, quick quality check of system. b) Early feedback on testing process.

**Risks:** Pre-acceptance test might discover many or high-priority errors at a pretty late stage (short before delivery). To avoid this, pre-acceptance test has to be scheduled considering this. If technically possible, a pre-acceptance test can already be executed before completeness of the system test.

**GSD specific characteristic:** a) Off-team as additional provider that you might not be familiar with yet.

b) Geographic distance leads to reduced informal ways of control which is necessary to monitor the test status. c) Possibly, the test environment is closer to later productive environment at On-site than at Off-site because of availability.

**Literature link:** In the literature a similar solution as the pre-acceptance test is described [41]. They propose Interim functional delivery by the Off-team with the goal that the On-team (TA) can execute a quality check early. Their intention is mainly to address the risk that heavy quality issues occur late in the project, i.e., shortly before delivery. We describe this in the risk of the pattern.

**Sources:** The pattern *Light-weight pre-acceptance test* was applied in several projects. One project mainly used offshoring in the second release, after onshore participation of off-colleagues in the first release. The team consists of eight Indian team members and one onshore PM who also took on the role of On-tester. The On-tester explained "The system test was in the responsibility of the Off-team. There the BA took on the role of tester. After system test the BA delivered the system to me for acceptance. Then I executed test cases of which I thought, knowing the specification in detail, that it (the system) could get stuck".

-------------------

In [42] we characterized the threats to validity of this work using a classification scheme described in [15]. Of course, most severe one is the threat to external validity: Only projects of one company, namely Capgemini are investigated. Thus, our findings might not hold in other contexts. However we took various measures to reduce the threat to external validity including a literature as a complementary source for our research.

## 5. Conclusions

With our solution of a collection of patterns for testing in GSD we reached several goals. 1) We close a gap in the literature, where so far little has been reported on the challenges and solutions for testing in GSD. 2) We provide practitioners at Capgemini sd\&m with easy to use solutions in the form of patterns when realizing test in GSD projects. With the problem/solution summary the user gets an overview. 3) Further, views of the patterns support the selection of patterns relevant for a certain role and test activity. The user benefits from the option to combine several patterns to set-up a tailored solution. We provide a basis in the investigation on the topic of testing in GSD. The identified problems and proposed solutions are a first step. A first review of the patterns by experienced test managers in offshore projects indicates the relevance and applicability of the patterns. Nevertheless, more investigation is needed to comprehensively analyze problems and solutions. Our research approach can serve as procedure for further investigation.

Future work contains the evaluation of the results in other projects within Capgemini sd\&m. This step is part of the current research and will be realized soon. As extension of the present work, an evaluation of the usage of the patterns in other companies would provide information if the results can be generalized. Findings from such evaluations should be introduced into the collection of patterns, e.g., modifications and/or extension of patterns or the description of further patterns.